\documentclass[useAMS,usegraphicx,usenatbib]{mn2e}
\usepackage{graphicx}
\usepackage{color}
\usepackage{amssymb,amsmath}

\def\lesssim{\mathrel{\hbox{\rlap{\hbox{\lower4pt\hbox{$\sim$}}}\hbox{$<$}}}}
\def\gtrsim{\mathrel{\hbox{\rlap{\hbox{\lower4pt\hbox{$\sim$}}}\hbox{$>$}}}}

\newcommand{\nside}{N_{\rm side}}

\title{Disks in the sky: A reassessment of the WMAP ``cold spot''}
\author[R. Zhang and D. Huterer]
{Ray Zhang$^{1}$\thanks{E-mail: rayzhang@umich.edu} and
  Dragan Huterer$^{1}$\thanks{E-mail: huterer@umich.edu}\\
  $^{1}$Department of Physics, University of Michigan, 450 Church St, 
  Ann Arbor, MI 48109}

\begin{document}

%\date{Accepted xxxx. Received xxxx; in original form xxxx}

\pagerange{\pageref{firstpage}--\pageref{lastpage}} 
\pubyear{2008}

\maketitle

\label{firstpage}

\begin{abstract}
  We reassess the evidence that WMAP temperature maps contain a statistically
  significant ``cold spot'' by repeating the analysis using simple circular
  top-hat (disk) weights, as well as Gaussian weights of
  varying width. Contrary to previous results that used Spherical Mexican Hat
  Wavelets, we find no significant signal at any scale when we compare the
  coldest spot from our sky to ones from simulated Gaussian random, isotropic maps. We trace this apparent
  discrepancy to the fact that WMAP cold spot's temperature profile just
  happens to favor the particular profile given by the wavelet. Since randomly
  generated maps typically do not exhibit this coincidence, we conclude that
  the original cold spot significance originated at least partly due to a
  fortuitous choice of using a particular basis of weight
  functions. We also examine significance of a more general
    measure that returns the most significant result among several choices of
    the weighting function, angular scale of the spot, and the statistics
    applied, and again find a null result.
\end{abstract}


\section*{Introduction}

Cosmic microwave background (CMB) maps have been studied in detail during the
last few years. These studies have been motivated by the remarkable full-sky
high-resolution maps obtained by WMAP \citep{Bennett2003,Spergel:2006}, and
led to a variety of interesting and unexpected findings.  Notably, various
anomalies have been claimed pertaining to the alignment of largest modes in
the CMB
\citep{TOH,deOliveira2004,Copi2004,Schwarz2004,Land2005a,lowl2,Abramo_align},
the missing power on large angular scales
\citep{Spergel2003,wmap123,Copi_nolarge}, and the asymmetries in the
distribution of power
\citep{Eriksen_asym,Hansen_asym,Bernui:2006ft,Hajian:2007pi}.  In the future,
temperature maps obtained by the Planck experiment, and large-scale
polarization information \citep{Dvorkin} may be key to determining the nature
of the large-scale anomalies.  For a review of the anomalies and attempts to
explain them, see \citet{Huterer_review}.

Several years ago, \citet{Vielva2004} reported an anomalously cold spot in the
WMAP microwave signal: kurtosis of the distribution of spots (defined using
Spherical Mexican Hat Wavelet weight functions) is unusually large on scales
of about $5\degr$, at $<0.5\%$ significance. The authors also noted that the
result is driven by a a cold spot in the southern hemisphere, at $(\ell,
b)=(-57\degr, 209\degr)$. The finding has been confirmed and further
investigated in \citet{Cruz2006}, who found that an equally cold or colder
spot of this size is expected in $<1\%$ of Gaussian random, isotropic skies,
as well as
\citet{Mukherjee2004,Cruz2005,McEwen2005,Cayon2005,Cruz2006,Rath,Naselsky2007,Pietrobon,Rossmanith},
some of whom also studied the spot's morphology.  The plot further thickened
when \citet{Rudnick} claimed that there is a corresponding cold spot
(underdensity in galaxy counts) --- in the NVSS radio survey, and at roughly
the same location as the CMB cold spot; however, this particular claim was
shown by \citet{Smith_Huterer} to be an artifact of the {\it a posteriori}
statistics and the particular way NVSS data had been analyzed. Nevertheless,
the CMB cold spot remains a much-studied topic and the source of
investigations of whether exotic physics could be the cause.

Perhaps surprisingly, nearly all of the works so far considered searches for
the cold spot using the same basis functions --- Spherical Mexican Hat
Wavelets (though with a few exceptions --- \citet{Pietrobon} used needlets,
while \citet{Rath} and \citet{Rossmanith} used the scaling indices). The only
variation in the different analyses was in the choice of the statistics that
was applied to the wavelet-based weights.

Here we set out to check the evidence for the cold spot using different, and
arguably simpler, set of weight functions.  We reassess evidence for the
``cold spot'' using circular top-hat weights (i.e.\ disks) of
arbitrary radius $R$. We do so in order to verify findings that relied on
wavelets, and more generally to investigate the robustness of the signal. We
also check results using simple Gaussian weights, finding results consistent
with those with the disks. We then investigate the source of this apparent
discrepancy with all of the previous work that used wavelets, and find that
the cause of the discrepancy is the specific temperature profile of the cold
spot which just happens to favor the profile of the Spherical Mexican Hat
Wavelet. In addition to the choice of the spots' weight function, the original
claim refers to the angular size of the spot of $\sim 5\degr$ that is also
chosen {\it a posteriori}. We investigate the effect of these choices by
defining a ``superstatistic'' measure that combines several previously
considered statistical measures of coldness and the associated choices of the
spot size and weight functions, and find that the claimed spot (or any other
spot in our sky) is not unusually significant using this new measure.

%DH Won't work because of section* (doesn't write out number of a section)
%The paper is organized as follows. We first define the weights we apply to
%calculate the spot temperatures, and describe the maps, in
%Sec.~\ref{sec:proc}. In Sec.~\ref{sec:results} we present results and describe
%them intuitively, and also test their robustness. In the same section, we
%discuss the results.  We conclude in Sec.~\ref{sec:conclude}.

\section*{Statistics and maps}\label{sec:proc}

\subsection*{Weight statistics}

The top-hat weights are familiar from structure formation (where they are
used in the definition of the amplitude of mass fluctuations over some scale
$R$, for example) and effectively represent another statistic to study the
cold spot.  We define the disk top-hat weight of radius $R$ as
\begin{equation}
D(r) \equiv A_{\rm disk}(R) \left [ \Theta(r)-\Theta(r-R)\right ],
\end{equation}
where $\Theta(x)$ is a Heaviside step function and $A_{\rm disk}(R)=\left
  (2\pi(1-\cos(R))\right )^{-1/2}$ is defined so that
\begin{equation}
\int_0^\pi D(r)^2 d\Omega=1.
\end{equation}
Note however that the normalization $A_{\rm disk}(R)$ is unimportant for
finding the coldest spot since we only do relative comparisons of temperatures
in disks on the sky.  The top hat-weighted temperature coefficients are given
by
\begin {equation}
  T_{\rm disk}(\hat{r}; R)=\int d\Omega' T(\hat{r}') D(\alpha; R),
\label{eq:T_disk}
\end{equation}
where $\hat{r}=(\theta, \phi)$ is the location of a given spot,
$\hat{r}'=(\theta', \phi')$ is the dummy location on the sky whose
temperature we integrate over, and $\alpha = \arccos(\hat{r} \cdot
\hat{r}')$ is the angle between the two directions.

The Gaussian weights that we use are defined equivalently. The weight
functions are 
\begin{equation}
G(r) \equiv A_{\rm Gauss}(R) \exp\left (-4\ln 2{r^2\over R^2}\right ),
\end{equation}
so that the full width at half maximum of the distribution is equal to
$R$. The weighted temperatures are given by
\begin {equation}
  T_{\rm Gauss}(\hat{r}; R)=\int d\Omega' T(\hat{r}') G(\alpha; R).
\label{eq:T_Gauss}
\end{equation}

Finally, the corresponding procedure applied to the wavelets is as follows
\citep{Cayon2001,MartinezGonzalez2002}. The Spherical Mexican Hat wavelets are defined as
\begin{equation}
  \Psi(\theta;R) = A_{\rm wav}(R)
\left (1+\left ({y \over 2}\right )^2\right )^2
\left (2-\left ({y \over R}\right )^2\right )
\exp{\left (-y^2 \over 2R^2\right )},
\end{equation}
where $y \equiv 2 \tan (\theta/2)$ and
\begin{equation}
A_{\rm wav}(R) = \left [2\pi R^2 \left (1+ {R^2 \over 2} + {R^4 \over 4}\right )\right ]^{-1/2},
\end{equation}
so that $\int d\Omega\, \Psi^2(\theta;R)=1$ over the whole sky. 
We can now define the continuous wavelet transform stereographically projected
over the sphere with respect to $\Psi(\theta;R)$, with $T$ being the CMB temperature:
\begin {equation}
  T_{\rm wav}(\vec{x}; R)=\int d\Omega' T(\vec{x}+\vec{\mu}') \Psi(\theta';R),
\end{equation}
where $\vec{x}\rightarrow (\theta,\phi)$ and $\vec{\mu'}\rightarrow
(\theta',\phi')$ are the stereographic projections to the sphere of the center
of the spot and the dummy location, respectively, and are given by
\begin{eqnarray}
  \vec{x} &=& 2\tan{\theta \over 2}(\cos \phi,\sin\phi),\\[0.1cm]
  \vec{\mu}' &=& 2\tan {\theta' \over 2} (\cos \phi',\sin\phi');
\end{eqnarray}
see \citet{MartinezGonzalez2002} for details.  To work in terms of purely
spherical coordinates, we center the spot location to the north pole of the
sphere, and rewrite the above as
\begin{equation}
  T_{\rm wav}(\hat{r}; R)=\int d\Omega' T(\hat{r}') \Psi(\alpha;R),
\label{eq:T_wav}
\end{equation}
where $M(\hat{r}')$ is the mask, defined to be 1 for pixels within the mask
and and 0 for those outside of it.  As the wavelet is effectively zero for
$\alpha$ values greater than $\sim 4$ times the radius, we can carry out the
integral by using the Healpix command {\tt query\_disc} to find all pixels
within a circle of that radius from the wavelet center.

To account for the masked parts of the sky, at each spot location $\hat{r}$ we
first calculate the ``occupancy fraction''
\begin{equation}
N(\hat{r}; R)=\int d\Omega' M(\hat{r}') \Psi^2(\theta;R).
\label{eq:occupancy}
\end{equation}	
We only include results for spot locations $\hat{r}$ for which
$N(\hat{r};R)>0.95$. Additionally, we do not include individual {\it pixels}
that have $M(\hat{r}')<0.9$ in order to limit biases due to masking (partially
masked pixels come about after degrading maps to a lower resolution). 
%There is remaining irreducible bias
  %induced by the few masked pixels that over- or underestimate of the overall
 % spot temperature. 
As discussed further below, we tested our procedures by using a higher occupancy
fraction and found consistent results.

\subsection*{Maps}
\begin{figure*}
\includegraphics[scale=0.30]{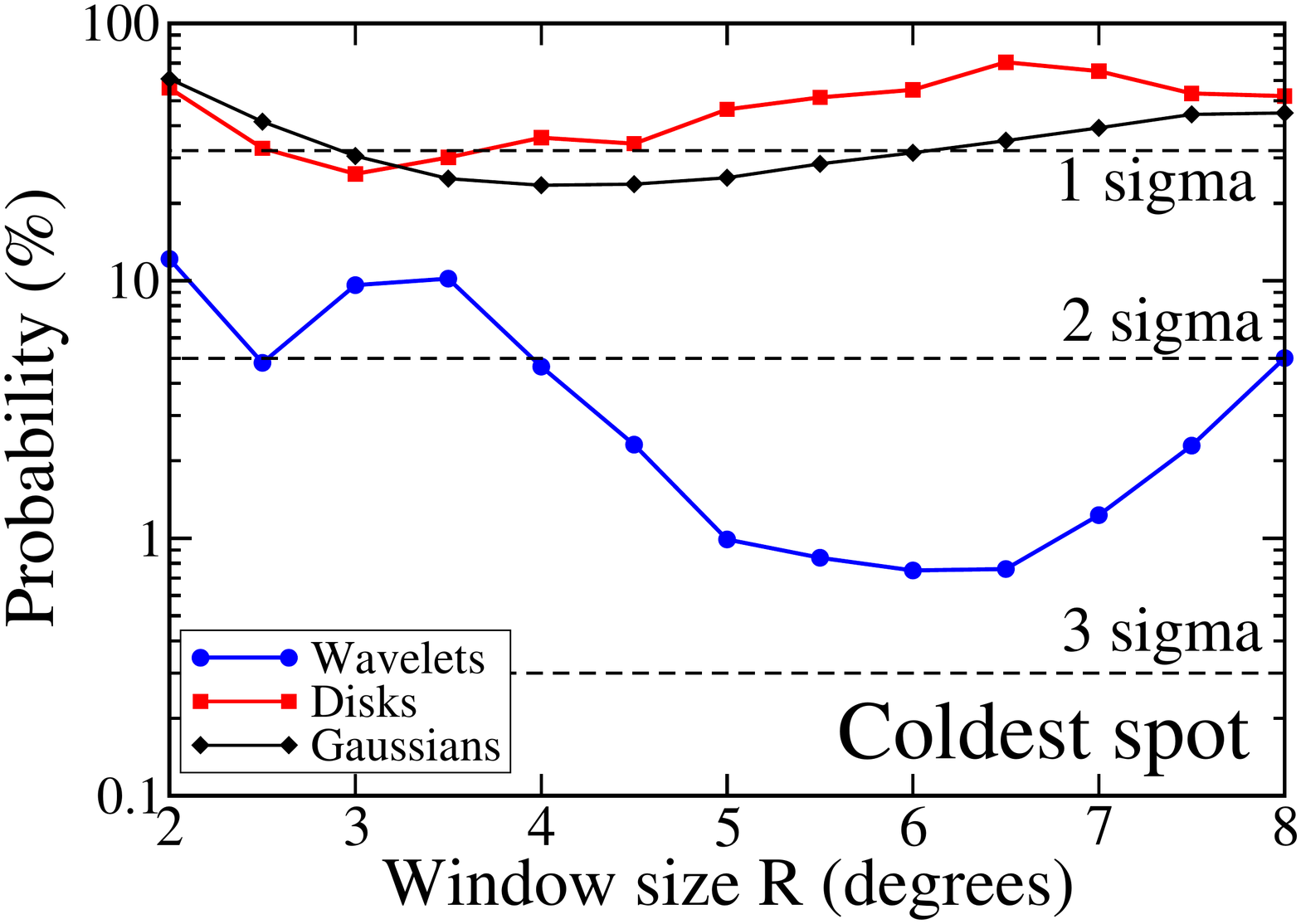}
\includegraphics[scale=0.30]{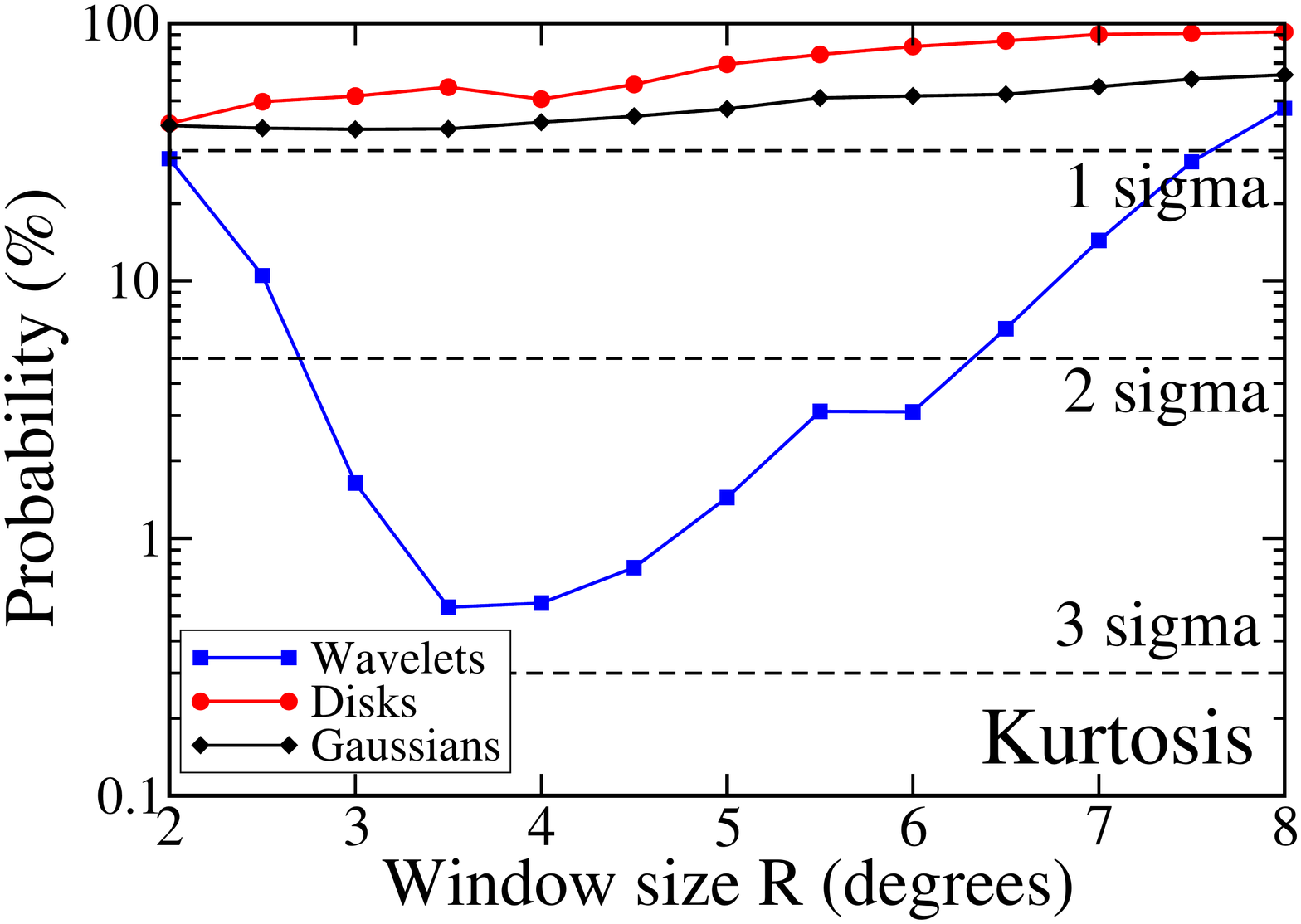}
\caption{ Significance in ``sigmas'' of the cold spot temperature $S^{\rm
    coldest}(R)$ (left panel) and kurtosis of the distribution of $S(R)$
  (right panel) for the three choices of weights that we have examined:
 disks (black), Gaussians (blue) and wavelets (yellow). Here $"1\sigma"$
  means $<32\%$ likely, $"2\sigma"$ means $<5\%$ likely, and $"3\sigma"$
  corresponds to $<0.3\%$ . Note that the scale $R$, shown on the x-axis, is
  defined separately for each choice of the weighting function so that the
  corresponding fair comparison can be made; see text for details. The results
  show that only the wavelet-based temperature cold spot and kurtosis deviate
  significantly from the Gaussian random expectation, and that the disk
  or Gaussian ones do not.}
\label{fig:signif}
\end{figure*}

We use WMAP's five year maps in our analysis \citep{WMAP5}. Following
\citet{Vielva2004}, the fiducial map we use is the coadded foreground-cleaned
map 
%Ray: Tried to clarify that our main analysis was on the foreground-cleaned
% combined Q-V-W map. This was because the Cruz paper used it.
\begin{equation}
	T = {\sum_{r=3}^{10}T_{r}(i)w_{r}(i)\over \sum_{r=3}^{10}w_{r}(i)},
\end{equation}
where $T$ is the coadded temperature, determined from the weighted sum of
temperatures $T_{r}$ of each individual radiometer $r\in
\{Q1,Q2,V1,V2,W1,W2,W3, W4\}$, divided by the total weight.  The weights at
each pixel for each radiometer are $w_{r}(i) = N_{r}(i) / \sigma_{r}^2$, where
$N_{r}(i)$ are the number of effective observations at the pixel, and
$\sigma_{r}$ is the noise dispersion for the given receiver.

This coadding was performed on maps at resolution of $\nside=512$ ($\sim 8'$),
then the KQ75 mask was applied.  As mentioned earlier, spots with more than
5\% of the weighted area masked ($N(\hat{r}; R)>0.95$) were not used.

The locations of centroids of spots are chosen to be centers of pixels in
$\nside=32$ resolution; therefore, we examine $N_{\rm pix}=12\,\nside^2\sim
12,000$ spots on the sky.  In order to calculate the spots' weighted
temperatures, however, we analyze the coadded map at the $\nside=128$ ($\sim
0.5\degr$) resolution, which is sufficiently high to lead to converged results
for $R\gtrsim 2\degr$ spots, yet sufficiently low to be numerically feasible.

The results of our analysis were then compared to 10,000 randomly generated
Gaussian full sky maps, with the same methodology applied. The skies have been
generated using the Healpix facility {\tt synfast}, and used as input the
power spectrum determined in the WMAP 5-year analysis \citep{Nolta_WMAP5}.
The maps were then smoothed by a Gaussian with FWHM $=1\degr$ to match
  the WMAP procedure.

\subsection*{Significance statistics}

The principal statistic that we use is the temperature of the coldest spot
divided by the standard deviation of the distribution of all spots
\begin{equation}
S_{\rm disk}(\hat{r}; R) \equiv {T^{\rm coldest}_{\rm disk}(\hat{r}, R)\over \sigma_{\rm
    disk}(R)}
\label{eq:S_stat}
\end{equation}
and equivalently for the Gaussian weights and the wavelets. Here $\sigma_{\rm
  disk}(R)$ is the standard deviation of the distribution of all spots in a
given map, while $T^{\rm coldest}_{\rm disk}(R)$ is the coldest spot in the
distribution. Note that the distribution of spot temperatures is not Gaussian
as we noted earlier, but this is irrelevant for us; we scale $T$ by $\sigma$
in Eq.~(\ref{eq:S_stat}) in order to account for small ($\sim 10\%$)
differences in the overall level of power in spots of characteristic size $R$
in the different maps --- in effect, $\sigma_{\rm disk}(R)$ provides units in
which to best report the coldest temperature. 

%In practice, scaling by $\sigma_{\rm disk}(R)$ (also done in \citet{Cruz2006})
%does not lead to results very different from those that would be obtained with
%reporting pure temperatures
%\dragan{Check}. 
%\ray{
% Actually, it does in some cases - the main reason I used it, I think, was
% that the standard deviations of our map was significantly different from
% that of random maps in disks (and also in Gaussians), so I wanted a fair
% basis for comparison.}
%Dragan: OK

Computing the significance of our statistic $S_{\rm disk}(\hat{r}; R)$ is then
in principle straightforward: we compare it to values obtained from simulated Gaussian
random maps and rank-order it; the rank gives the probability.

In addition to the cold spot significance, we follow \citet{Vielva2004} and
\citet{Cruz2005} and consider the kurtosis of spots in a given map. The
kurtosis is simply related to the fourth moment of the distribution of the
spots
\begin{equation}
K_{\rm disk}(R) \equiv {1\over N_{\rm spots}}
{\sum_{i=1}^{N_{\rm spots}}T_{\rm disk}(\hat{r_i}, R)^4\over 
\sigma_{\rm disk}(R)^4}-3
\label{eq:kurtosis}
\end{equation}
and equivalently for the Gaussian weights and the wavelets.

\section*{Results}\label{sec:results}

\begin{figure*}
\includegraphics[scale=0.30]{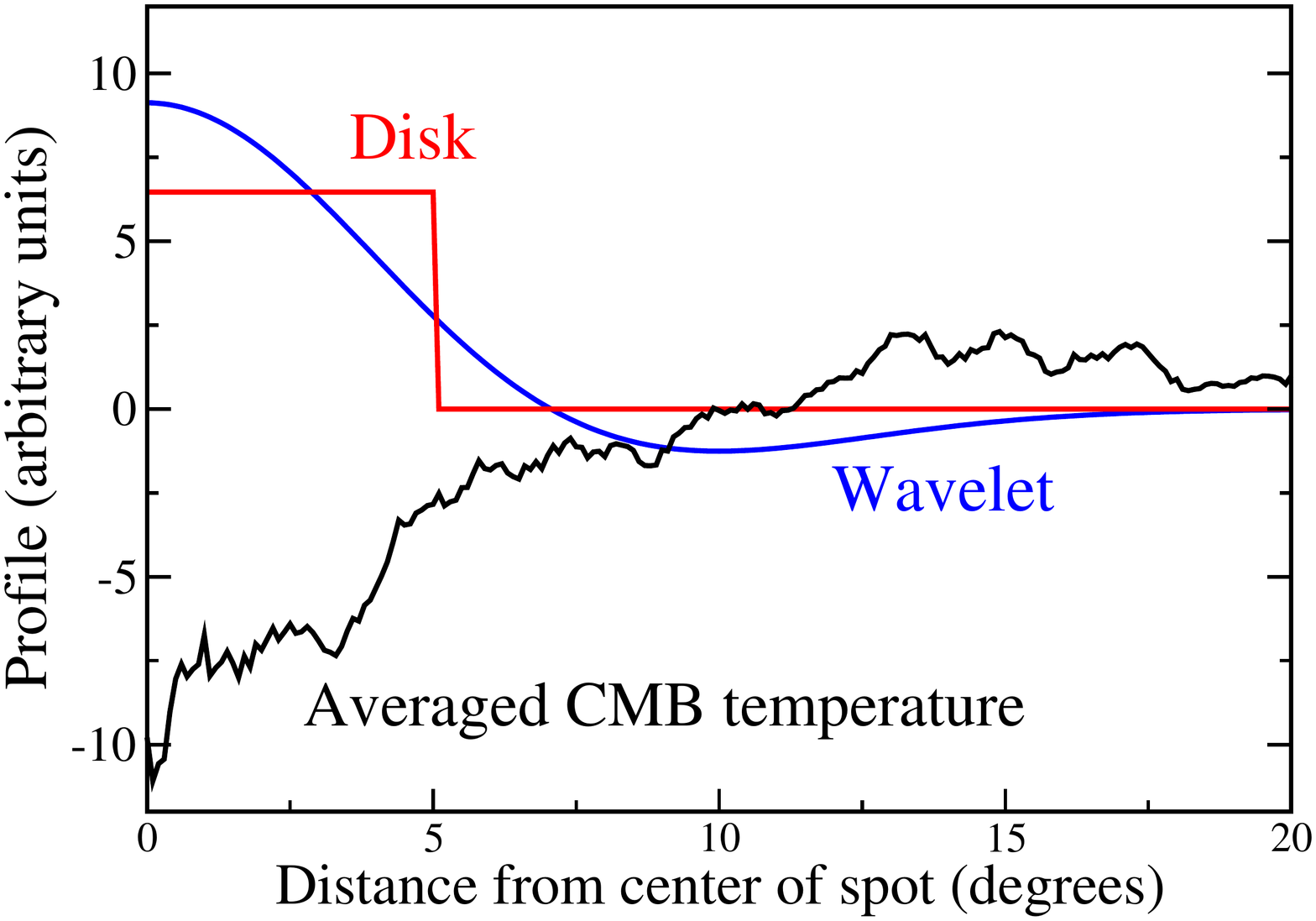}
\includegraphics[scale=0.30]{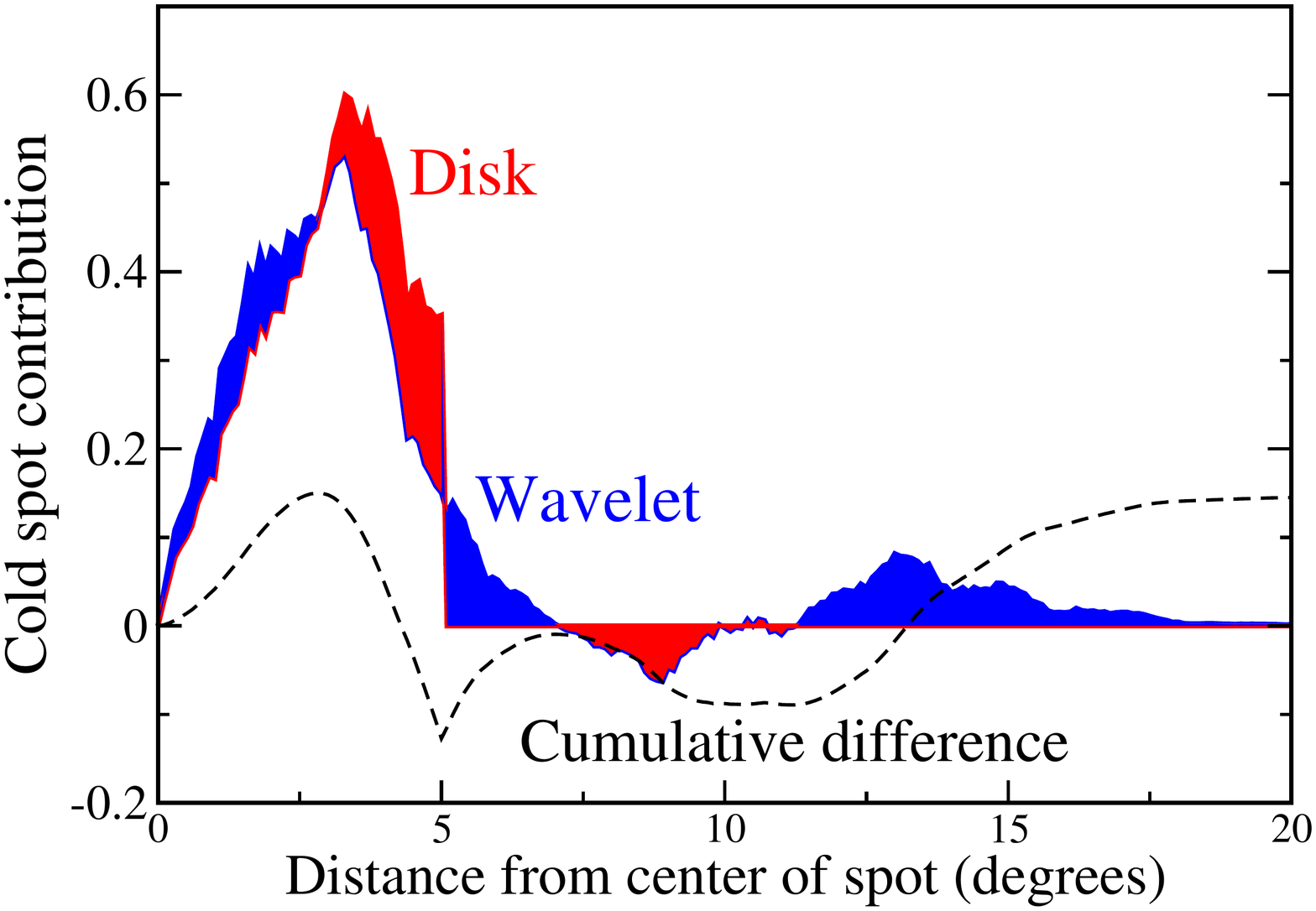}
\caption{Left panel: disk and wavelet weights, together with the
  azimuthally-averaged CMB temperature profile as a function of radial
  distance from the (wavelet-found) center of the cold spot. Right panel: contribution to
  the weighted temperature using disks and wavelets; the blue (red) shaded
  areas show the difference by which wavelets (disks) dominate in the given
  ranges of distance, while the dashed line shows the cumulative difference
  between the two. Note that all lines in both panels have arbitrary
  normalization, although the disk and wavelet lines are consistently
  compared using their fiducial normalizations from the text. }
\label{fig:profile}
\end{figure*}

\subsection*{Wavelet weighted spot}

We first make sure that we reproduce the cold spot results of
\citet{Vielva2004} and \citet{Cruz2005}.  For Spherical Mexican Hat Wavelets
with $R=5\degr$, we find the center of the coldest spot in the five-year
combined cleaned map, is at coordinates $(\ell, b)=(-57.7\degr, 209.3\degr)$
(corresponding to spherical coordinates $(\theta, \phi)=(147.7\degr,
209.3\degr)$).  In general agreement with the \citet{Cruz2007} results, we
find that only $(0.99\pm 0.10)\%$ of simulated statistically isotropic,
Gaussian random maps exhibit a more significant cold spot (i.e.\ a more
negative value of $S^{\rm coldest}_{\rm wav}(R)$) for this value of
$R$. Here and throughout, the error bars account for
  the finite number of ($N=10,000$) simulated maps; we quote the standard margin
  of error which, for a fraction $p$ of a total of $N$ events, is given as
  $\sigma(p)=\sqrt{p(1-p)/N}$. Moreover, we confirm that while the variance
and skewness of the distribution of $S^{\rm coldest}_{\rm wav}(R)$ from
synthetic maps, the kurtosis at $R= 5\degr$ is high at the $(1.44 \pm 0.12)\%$
confidence.

Reporting the significance only for the $R=5\degr$ may be unfair,
however. To address this, we consider the range $2\degr\leq R\leq 8\degr$ in
steps of $0.5\degr$ (the lower bound is set to correspond to spots
significantly larger than smoothing of the maps of roughly $1\degr$).  The
significances of these results are given as black lines in the two panels of
Fig.~\ref{fig:signif}\footnote{These tests are
    computing-intensive, and we were forced to compare to simulated maps at
    $\nside=32$ (rather than $\nside=128$) for wavelets for $6.5\degr\leq R\leq 8\degr$; we
    have checked at lower values of $R$ that the results at the two
    resolutions are in good agreement.}.  The significance of the
wavelet-determined cold spot peaks around $R=5$-$6\degr$, while the kurtosis is
significant in the range $3\degr\leq R\leq 5\degr$.

\subsection*{Disk and Gaussian weighted spot}

We now repeat the same analysis with the circular weights. First we confirm
that the coldest disk-weighted spot in the co-added smoothed Q-V-W map is at
nearly the same location as the wavelet-weighted spot, at $(\ell, b)=(-57.4,
208.0)$. The result is similar in the foreground cleaned Q-V-W map.  However,
the disk-weighted spot is not unusually cold: at all scales $R$ between
between $2\degr$ and $8\degr$, the statistic $S^{\rm coldest}_{\rm disk}(R)$
is not unusually low relative to expectations from Gaussian random maps; see
the left panel of Fig.~\ref{fig:signif}. We find the same results for the
kurtosis of the distribution of $S_{\rm disk}(\hat{r}; R)$ --- as shown in the
right panel of Fig.~\ref{fig:signif} these distributions fall well within the
expectation on all scales we examined.
%skewness is lower than expected, but not too significant at a minimum of 4.94\%; 

Surprised by these results, we have repeated the same tests with Gaussian
weighting, where the full-width half-maximum (FWHM) of the Gaussian weight has
been set equal to the scale $R$; this way we ensure that a large fraction
(about $94\%$) of the weight is applied within the radius $R$. The results are
similar as for the disks in that they are not significant; see again
Fig.~\ref{fig:signif}. In particular, the coldest spot is most significant at
$R=4$-$5\degr$, but even there only at the 1-$\sigma$ ($25\%$) level, while
the kurtosis is not significantly large or small at any scale.

\subsection*{The density profile of the cold spot}

The question is obvious: Why was the cold spot so significant for wavelets,
but not so much for disks? To address this, we show the disk and wavelet
weights, together with the azimuthally-averaged CMB temperature profile, as a
function of radial distance from the (wavelet-found) center of the cold spot
in the left panel of Fig.~\ref{fig:profile}. In the right panel, we show
contribution to the weighted temperature using disks and wavelets, as well
as the cumulative difference between the two. This figure shows the case of
$R=5\degr$ which approximately maximizes significance of the wavelet-based cold spot. 

The azimuthally averaged density profile of the temperature is about
equally distributed between zero and $5\degr$ (that is, the blue area in the
right panel is about the same as the red one up to $5\degr$). However, at
distance beyond the edge of the disk of $5\degr$, the wavelet
accumulates more weight as seen in the right panel. The reason is shown with
the curve labeled ``averaged CMB temperature'' in the left panel: the CMB
profile goes from negative to positive with increasing radius from the center
of the (wavelet-based) cold spot, precisely favoring the wavelet profile that
has roughly the opposite behavior. 

What is the likelihood of this conspiracy that the temperature profile of the
coldest spot mimics the shape of the wavelet?  Using Gaussian random maps, we
estimate the likelihood that a given map has the wavelet-determined cold spot
is more significant than the disk-determined cold spot by {\it at least} as
much as in WMAP where $|S^{\rm coldest}_{\rm wav}(R)-S^{\rm coldest}_{\rm circ
}| = 1.5$. We find that the wavelets are more significant that
  the disks by at least this margin in only $(1.89 \pm 0.13) \%$ of the random
  maps (while the disks are as or more significant in only $(1.96 \pm 0.14)
  \%$  cases). From this, we conclude that typical Gaussian random CMB maps do
not show increased significance of the wavelet determined cold spot, relative
to the disk-determined one, to the same extent as our sky does.

\begin{table*}
\begin{tabular}{|l|l|l|l||l|l|l||}
\hline\hline
& \multicolumn{3}{c}{Cold spot statistic ($R=5\degr$)} 
& \multicolumn{3}{c}{Kurtosis statistic  ($R=5\degr$)}\\\hline\hline         
                      & \rule[-2mm]{0mm}{6mm} Wavelet 
                      & \rule[-2mm]{0mm}{6mm} Disk 
                      & \rule[-2mm]{0mm}{6mm} Gaussian
                      & \rule[-2mm]{0mm}{6mm} Wavelet 
                      & \rule[-2mm]{0mm}{6mm} Disk
                      & \rule[-2mm]{0mm}{6mm} Gaussian
\\\hline\hline
\rule[-2mm]{0mm}{6mm} Value of statistic \qquad\qquad
                      & -3.21 & -4.54 & -3.52
                      & 0.58 & -0.23 & -0.06
                      \\\hline
\rule[-2mm]{0mm}{6mm} Significance in \%    
%                      & $0.99\pm 0.10$ & $46.27\pm 0.50$  & $25.11\pm 0.43$ 
%                      & $1.44\pm 0.12$ & $69.26\pm 0.46$  & $46.52\pm 0.50$ 
                      & $0.99\pm 0.10$ & $46.3\pm 0.5$  & $25.1\pm 0.4$ \qquad\qquad
                      & $1.44\pm 0.12$ & $69.3\pm 0.5$  & $46.5\pm 0.5$ 
                      \\\hline\hline
\end{tabular}
\label{tab:table}
\caption{Statistics and its significance for coldest spot and kurtosis of
  spots, evaluated on scale $R=5\degr$, for the three weights we considered. 
  For the coldest spot, the statistics is $S$, defined in
  Eq.~(\ref{eq:S_stat}), while for kurtosis, the statistic is just its value, defined 
  in Eq.~(\ref{eq:kurtosis}). The error bars are Poisson and reflect the
  finite number of the simulated Gaussian random maps to which we compared
  WMAP. The results are robust to variation in the
  choice of the WMAP map, or radius $R$, as discussed in the text.  }
\end{table*}

\subsection*{Robustness}

We have tried varying a number of details, with the following results:

\begin{itemize}
\item In addition to the coadded {\it foreground cleaned} Q+V+W map, we also
  used the coadded Q+V+W map, the coadded V+W map, and the coadded foreground
  cleaned V+W maps available from WMAP. All of the
    significances are very close to the foreground QVW map value; we have
    checked that smaller scales ($1$-$1.5 \degr$), which were not in the range
    we presented in the final analysis, would be somewhat discrepant; this is
    not surprising given that the maps are smoothed to $1\degr$.

\item To test the effects of the resolution of the map, we vary the resolution
  from $\nside=128$ ($\sim 0.5\degr$ pixels) to $\nside=32$ ($\sim 2\degr$
  pixels), with 16 times fewer pixels. We again find
  consistent results except at small scales, $R<2\degr$, which makes sense
  since pixelization is expected to play a role only when pixel size becomes
  comparable to the spot scale $R$.

\item To ensure that our finite step size has not accidentally overlooked a
  cold spot, we refine the resolution in our search for the coldest spot in
  WMAP by querying at {\it every} pixel in an $\nside=512$ map within $3\degr$
  of the center of the reported cold spot; this stepping size is effectively
  $\sim 256$ times higher than before.  While an increase in the temperature
  of the cold spot is entirely expected, we find that this increase is small
  enough not to appreciably change the significance results for all three choices of the
  weight function.

\item To test our prescription for dealing with partially masked spots, where
  we only analyze spots that have the ``occupancy fraction'' $N(\hat{r}; R)>
  0.95$ (see Eq.~(\ref{eq:occupancy})), we repeat the analysis with the
  minimum occupancy number of $0.98$. While the resulting {\it number} of
  spots retained in the analysis is now much smaller, decreasing by between tens
  of percent (for spots at $R=2\degr$) to about a factor of 10 (for $R=8\degr$
  disks), we find results generally consistent with our fiducial case:
  the statistic $S$ and kurtosis calculated using the wavelet weights are significant, while the
  same statistics calculated using the disk and Gaussian weights are not.
\end{itemize}

\subsection*{More general tests}

It is clear that {\it a posteriori} choices were made in the original claims
for the existence of the cold spot --- in addition to the choice of the
weighting function (which is the principal subject of this paper), the moment
of the distribution of spots (kurtosis) and spot scale ($5\degr$) have been
called out {\it after} noticing that they are unusual\footnote{While
  observation of some of the other anomalies mentioned in the introduction was
  technically also {\it a posteriori}, those anomalies had to do with special
  scales (e.g.\ largest observable scales on the sky) or directions (e.g.\ the
  ecliptic, which the telescope pointings preferentially avoid). In contrast,
  there appears to be nothing special about kurtosis of the spot distribution,
  or scales of $\sim 5\degr$.}. In contrast, variance and skewness of the
spot distribution, or kurtosis and scales larger or smaller than $\sim
5\degr$, do not show departures from expectations based on Gaussian random
isotropic maps, as we have checked as well.

We have investigated how results change with more general tests as follows. We
have formed a ``superstatistic'' defined as maximum significance of either variance,
skewness, kurtosis, or coldness (the last two being defined earlier in this
section) over any scale $R$ or weight function set $W$
\begin{equation}
\mathcal{S}_{\rm super} \equiv \underset{\text{R, W, {\rm Stat}}}{\text{max}} 
\{ P({\rm Stat}_W(R))\}
%&&\left \{ P({\rm Var}_W(R)), P({\rm Skew}_W(R)), \right . \nonumber \\[-0.1cm]
%&& \left . P({\rm K}_W(R)),   P(S_W(R)) \right \}
\label{eq:superstatistic}
\end{equation}
where\footnote{For wavelets with $6.5\degr \leq R \leq 8\degr$ the
  calculations at $\nside=128$ were unfeasible due to the large number of
  pixels to keep track of.  While we have checked that $\nside=32$ wavelet
  results are similar to $\nside=128$ at $R<6.5\degr$, and show no significant
  results at larger scales, for consistency we decided to quote the
  superstatistic results at $\nside=128$ and consider the $2\degr \leq R \leq
  6.5\degr$ scales for the wavelets, and $2\degr \leq R \leq 8\degr$ for the
  disks and Gaussians.}

\begin{eqnarray}
R &\in& \{2\degr, 2.5\degr, \ldots, 8\degr \} \nonumber \\[0.1cm]
W &\in& \{{\rm wavelet}, {\rm disk}, {\rm Gaussian} \}\\[0.1cm]
{\rm Stat} &\in& \{{\rm Variance}, {\rm Skewness}, {\rm Kurtosis}, S
\}\nonumber
\label{eq:superstat_vary}
\end{eqnarray} 
and where each probability $P$ was individually calculated relative to
Gaussian random isotropic maps as described earlier. [The $S$ and kurtosis
statistics have been defined in Eqs.~(\ref{eq:S_stat}) and
(\ref{eq:kurtosis}), while the variance and skewness are defined analogously
to kurtosis.] Note that we define $P$ to capture the possibility that the
statistics in question is either small or large relative to expectation; in
other words, we adopt the minimum of $r$ and $(100\%-r)$, where $r$ is the rank
of the statistics relative to simulated maps.

\begin{figure}
\includegraphics[scale=0.30]{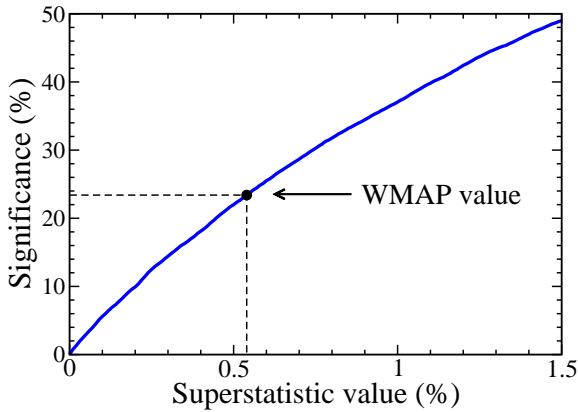}
\caption{Significance of the superstatistic $\mathcal{S}_{\rm
      super}$, defined in Eq.~(\ref{eq:superstatistic}), as a function of its
    value. The line shows results calibrated from our simulated maps, while
    the point denotes the value (and the unremarkable significance) from
    WMAP. }
\label{fig:superstatistic}
\end{figure}

We find that the value of the $\mathcal{S}_{\rm super}$ statistic for the WMAP
cleaned QVW map is 0.54\%, and this value is attained by the
kurtosis statistic $S_{\rm wavelet}$ at scale $R=3.5\degr$. However, this
value is not too unusual: we find that 23\% of the Gaussian random skies have
a smaller value of $\mathcal{S}_{\rm super}$; see
Fig.~\ref{fig:superstatistic}.  

To  test the robustness of this result, we consider an
  alternative, more restricted, definition of the superstatistic where only
  the wavelet weights are considered, but where we still varying scale $R$ and
  statistic Stat; see Eq.~(\ref{eq:superstat_vary}).  Here we effectively
  assume that, for whatever reason, wavelets are the preferred weight
  functions to be used, but we still seek to avoid the {\it a posteriori}
  choices of the scale and statistics.  The new superstatistic
%  value in our sky remains the same, and it 
  is again not statistically significant; we find that 15\% of Gaussian
  random skies show greater significance.

Thus, superstatistic results confirm our earlier conclusion that less
{\it a posteriori} tests do not indicate a statistically significant cold spot
in the WMAP data.

\section*{Conclusions}\label{sec:conclude}

The ``cold spot'', together with low power at large angles, multipole
alignments, north-south power asymmetry, has been one of the most studied
anomalies in WMAP CMB temperature maps. So far there have been
  no compelling proposals, cosmological or systematic, that would explain
  existence of the claimed cold spot, which is perhaps not surprising given
  that neither its radius ($\sim 5\degr$) nor its direction in the sky are
  particularly special.

In this paper, we have investigated evidence for the cold spot. While
we confirmed its high statistical significance using the wavelet basis of
weight functions, we did not confirm the existence using the disk top-hat,
or Gaussian, weights. The cold spot is indeed at the same location in WMAP
maps with the latter two bases, but it is not significant when compared to
expectation based on Gaussian random, isotropic skies. 

We traced the apparent inconsistency to the fact that the radial temperature
profile around the cold spot center is such that it favors the wavelet
profile; see Fig.~\ref{fig:profile}. This is a chance event, since only 5\% of
the Gaussian random, isotropic skies exhibit equal or more significant
discrepancy in favor of the wavelets. Moreover, we found that the result is
insensitive to the choice of the map or the statistic used for the cold
spot. 

Motivated by these findings, we also examined significance of a more general
measure -- which we called the ``superstatistic'' -- that combines the various
choices of the weighting function, spot size, and statistics, and returns the
most significant choice consistently for each map. We again find a null
result; the WMAP superstatistic is low only at $\sim 20\%$ level relative to Gaussian
random and isotropic simulated maps.

Therefore, we find no compelling evidence for the anomalously cold spot in
WMAP at scales between 2 and 8 degrees. The existing evidence
apparently hinges on the particular choice of the weight functions to define
the spot (Spherical Mexican Hat Wavelets) and their scale ($R\sim 5\degr$).
While our conclusion may sound like a depressing null result, we are upbeat
about future tests with WMAP (and soon, Planck) to uncover and test unexpected
features and anomalies.

\section*{Acknowledgements} 

We acknowledge use of the HEALPix \citep{healpix} package, and also the Legacy
Archive for Microwave Background Data Analysis (LAMBDA). We thank Christoph
R{\"a}th and Patricio Vielva for useful communications.  DH is supported by the
DOE OJI grant under contract DE-FG02-95ER40899, NSF under contract
AST-0807564, and NASA under contract NNX09AC89G. He thanks the Aspen Center
for Physics for hospitality while this work was nearing completion.

\bibliographystyle{mn2e}
\bibliography{wmap_coldspot}

\label{lastpage}
\end{document}